\documentstyle[prl,aps,epsf,multicol]{revtex}
\begin{document}

\title{The universal behavior of one-dimensional, multi-species
branching and annihilating random walks with exclusion}
\author{G\'eza \'Odor}
\address{Research Institute for Technical Physics and 
Materials Science, \\ H-1525 Budapest, P.O.Box 49, Hungary}    
\maketitle

\begin{abstract}
A directed percolation process with two symmetric particle 
species exhibiting exclusion in one dimension is investigated 
numerically. It is shown that if the species are coupled 
by branching ($A\to AB$, $B\to BA$) a continuous 
phase transition will appear at zero branching rate limit 
belonging to the same universality class as that of the 
dynamical two-offspring (2-BARW2) model. 
This class persists even if the branching is biased towards 
one of the species.
If the two systems are not coupled by branching but hard-core
interaction is allowed only the transition will occur at finite 
branching rate belonging to the usual $1+1$ dimensional directed
percolation class.
\end{abstract}
\pacs{\noindent PACS numbers: 05.70.Ln, 82.20.Wt.}

\begin{multicols}{2}

The study of phase transitions in low dimensions is an
interesting and widely investigated topic \cite{Dick-Mar,Hin2000} 
(since the mean-field solution is not valid).
The research of non-equilibrium phase transitions occurring in
one dimensional coupled systems has drawn interest nowadays
\cite{HT97,Odo00,HayeDP-ARW,Park,barw2cikk,Lipo,CoupledDP,uni-PC,dimercikk,Frei,Wij,Trimp}.
Several models have been found with transitions that do not 
belong the robust directed percolation (DP) class 
\cite{Jan81,Gras82,DP} or to the parity conserving (PC) class 
\cite{Gras84,ujMeOd} which are the most prominent ones among 
one component systems.
Particle blocking which is common in one dimension has not been 
taken into account in field theoretical description of these models
yet \cite{Cardy-Tauber,Janssen-col}.
It has been known for some time that the pair contact process
\cite{PCP} can be regarded as a coupled system that exhibits
DP class static exponents while the spreading ones 
depend on initial densities \cite{MDGL}.
The field theoretical investigation of Janssen \cite{Janssen-col} 
predicts that in coupled DP systems the symmetry between species 
is unstable and generally a phase transition belongs to the 
class of unidirectionally coupled DP where coupling between
pairs of species is relevant in one direction only. 
Such systems have been shown to describe also certain surface 
roughening processes \cite{CoupledDP,uni-PC}.

Recently we have shown \cite{arw2cikk} that in the two-component 
annihilating random walk ($AA\to\emptyset$, $BB\to\emptyset$)
owing to the hard-core interaction of particles dynamical
exponents are non-universal.
Some consequences of hard core effects for random walks in 
one dimension have been known for some time already \cite{Dhar}.

Very recently simulations \cite{Park,barw2cikk} gave numerical 
evidence that in the two-component branching and annihilating 
random walk (2-BARW2) the lack of particle exchange between 
different species results in new universality classes in 
contrast to widespread beliefs that bosonic field theory can 
well describe these systems. 
The critical exponents obtained numerically suggest that the 
location of offspring particles at branching is the relevant 
factor that determines the critical behavior. 
In particular if the parent separates the offsprings: 

1) $A\to BAB$ the steady state density will be
higher than in the case when they are created on
the same site: 

2) $A\to ABB$ for a given branching rate 
because in the former case they are unable to annihilate
with each other. This results in different order parameter
exponents for the symmetric (2-BARW2s) and the asymmetric
(2-BARW2a) cases ($\beta_s=1/2$ vs $\beta_a=2$
for 1) and 2), respectively).

Hard-core effects are conjectured to cause a series of 
new universality classes in one dimension \cite{Park}.
In this paper I point out that probably only a few
universality classes emerge as the consequence of
particle exclusion, because other symmetries and 
conservation laws (like that of the PC class) 
will become irrelevant.

In the present study first I show that in case of the 
two-component single off-spring BARW model (2-BARW1)
defined as
\begin{eqnarray}
A\stackrel{\sigma_A/2}{\longrightarrow} AB  \ \ \ \ , \ \ \
A\stackrel{\sigma_A/2}{\longrightarrow} BA \\
B\stackrel{\sigma_B/2}{\longrightarrow} BA  \ \ \ \ , \ \ \
B\stackrel{\sigma_B/2}{\longrightarrow} AB \\
AA\stackrel{\lambda}{\longrightarrow}\text{\O} \ \ \ , \ \ \
BB\stackrel{\lambda}{\longrightarrow}\text{\O} \\
A\text{\O}\stackrel{d}{\leftrightarrow}\text{\O}A \ \ \ , \ \ \
B\text{\O}\stackrel{d}{\leftrightarrow}\text{\O}B \\
AB\stackrel{0}{\leftrightarrow}BA \label{proc}
\end{eqnarray}
a continuous phase transition will occur at zero branching rate
limit ($\sigma=0$) like in the 2-BARW2 model where they are
equivalent and therefore the exponents on the critical point 
must be the same as those determined in \cite{Park,barw2cikk,arw2cikk}. 
Furthermore I show that the order parameter exponent describing 
the singular behavior of the steady state density near the critical 
point coincides with that of the 2-BARW2s model. 

\begin{figure}
\epsfxsize=70mm
\centerline{\epsffile{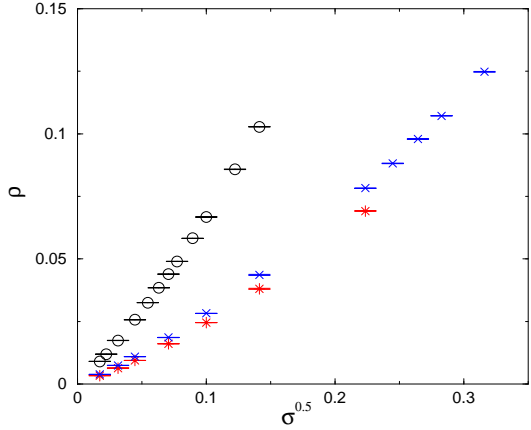}}
\vspace{2mm}
\caption{Steady state density as the function of $\sigma^{0.5}$
in the one-dimensional 2-BARW1 model. 
Circles correspond to $\rho_A+\rho_B$ for $\sigma_A=\sigma_B$;
crosses to $\rho_A$, stars to $\rho_B$ when $\sigma_A=\sigma_B/2$.
\label{dp2rho}
}
\end{figure}

\begin{figure}
\epsfxsize=70mm
\centerline{\epsffile{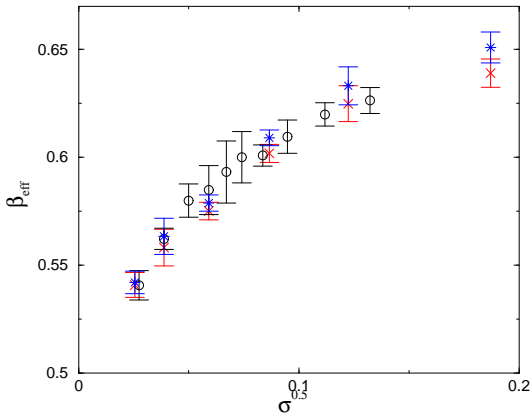}}
\vspace{2mm}
\caption{Effective $\beta$ in the 2-BARW1 model as the function
of $\sigma^{0.5}$. Different symbols denote the same as
in Fig.\ref{dp2rho}.
\label{betadp2}
}
\end{figure}

The particle system was simulated on a lattice with size 
$L=4\times10^4$ and periodic boundary conditions for 
different $\sigma$-s (with $\lambda=d=1-\sigma$ condition). 
The initial condition was uniformly random distribution of 
$A$-s and $B$-s with a total concentration
$0.5$. The evolution of the density was followed
until steady state has been reached plus $t\sim 10^4$ 
Monte Carlo sweeps (throughout the whole paper 
$t$ is measured in units of Monte Carlo sweeps 
(MCS) of the lattices). As Figure \ref{dp2rho} shows 
a phase transition occurs at $\sigma_A=\sigma_B=0$ indeed.
The order parameter exponent has been determined with
the local slope analysis of the data
\begin{equation}
\beta_{eff} (\sigma) = \frac {\ln \rho_{i} -\ln \rho_{i-1}}
              {\ln \sigma_i - \ln \sigma_{i-1}} \ \ ,
\end{equation}
providing an estimate for the true asymptotic behavior of the
order parameter
\begin{equation}
\beta = \lim_{\sigma\to 0} \beta_{eff}(\sigma) \,.
\end{equation}
As one see on Figure \ref{betadp2} $\beta_{eff}$ extrapolates
to $\beta=0.50(1)$ with a strong correction to scaling like
in case of the 2-BARW2s model \cite{barw2cikk}. 
The coincidence of this off-critical exponent in addition to 
the equivalence of processes at the critical point assures 
that they belong to the same universality class.
If we destroy the symmetry between species by the branching 
rates: $\sigma_A=\sigma_B/2$ we still get the same order parameter
exponents ($\beta=0.50(1)$) for both species (Fig.\ref{betadp2}).
Therefore this universality class is stable with respect to coupling
strengths unlike the coloured and flavoured directed percolation 
\cite{Janssen-col}. 

It is also insensitive whether or not the parity of particles is 
conserved meaning that the $A\to BAB$ process can be decomposed
to a sequence of $A\to AB$, $AB\to BAB$ processes. This may
seem to be quite obvious when particle exchange is not allowed, 
and if locality is assumed. By the choice of parameters 
$d=1-\sigma$ in the neighbourhood of the critical point the
the diffusion is strong and the locality condition is not
met. Still the two process share the same critical behavior. 

\begin{figure}
\epsfxsize=70mm
\centerline{\epsffile{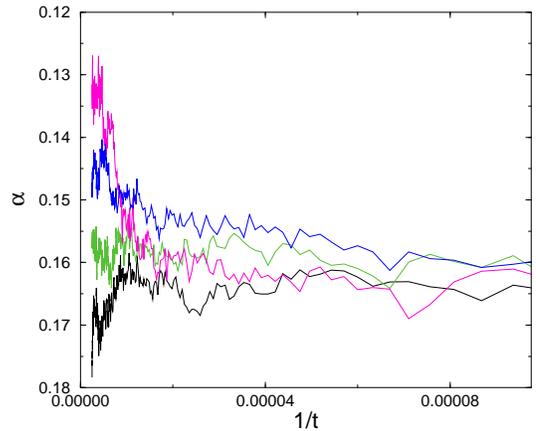}}
\vspace{2mm}
\caption{Effective decay exponent $\alpha_{eff}$ as the function of
$1/t$ in the decoupled two component DP model. The system size is
$L=4\times10^4$, the decay is followed for $2\times10^5$ MCS.
The different curves correspond to: 
$\sigma=0.81103$, $0.81107$, $0.81109$, $0.8111$
(from bottom to top).
\label{dp2s}
}
\end{figure}

If we decouple the two systems and allow hard-core exclusion only:
\begin{eqnarray}
A\stackrel{\sigma}{\longrightarrow} AA \\
B\stackrel{\sigma}{\longrightarrow} BB \\
AA\stackrel{\lambda}{\longrightarrow}\text{\O} \ \ \ , \ \ \
BB\stackrel{\lambda}{\longrightarrow}\text{\O} \\
A\text{\O}\stackrel{d}{\leftrightarrow}\text{\O}A \ \ \ , \ \ \
B\text{\O}\stackrel{d}{\leftrightarrow}\text{\O}B \\
AB\stackrel{0}{\leftrightarrow}BA \label{procu}
\end{eqnarray}
the critical point will be shifted to $\sigma=0.81107(1)$ and DP like
density decay can be observed on the local slopes defined as
\begin{equation}
\alpha_{eff}(t) = {- \ln \left[ \rho(t) / \rho(t/m) \right] 
\over \ln(m)} \label{slopes}
\end{equation}
(where we use $m=8$ usually) (see Figure\ref{dp2s}).
One can not observe any relevant correction to scaling here, the
most straight curve corresponding to the critical one 
($\sigma=0.81107$) extrapolates to $\alpha=0.158(2)$ which agrees
very well with the $\beta/\nu_{||}=0.159464(6)$ value of 
the $1+1$ DP class value that can be found in the
literature \cite{IJensen99}. This is different from the case
of coupled annihilating random walk, where the blocking 
causes marginal perturbation to the standard decay process 
\cite{arw2cikk}.

One can generalize the results by taking into account that
neighbouring $AA$ and $BB$ offsprings decay very quickly and
therefore irrelevant for the leading scaling behavior.

{\it Conjecture : In coupled, one dimensional N-component
BARW systems with particle exclusion and branching processes 
like: $A\to BABB$, $A\to BAAA$, $A\to BAC$ ... leaving behind 
non-reacting neighbouring particles which block each other 
the universality class of a phase transition 
will be the same as that of 1-BARW2s.}
{\it If the branching creates only pairs that can
annihilate immediately (like: $A\to BAAB$ ... etc.)
the class of transition will be the same as that of the 
2-BARW2a model.}
We can also conclude that in case of reaction-diffusion
processes where spontaneous decay is allowed:
$2A\to A$, $A\to\text{\O}$ the blocking effect between
dissimilar species is irrelevant.

It is very likely that the transition of a very recently 
introduced ladder model \cite{Lipo} also belongs to this class.
This model is composed of two one dimensional subsystems
following BARW at the critical point and coupled by ladder
links. In the supercritical region by updating an active
site one can create an offspring on the other subsystem 
or increase the inactivity level of that site. For small
coupling strength ($s=1$) the very few blocking events 
can not introduce relevant blocking on the other 
subsystem and the scaling exponents agree with those of
the coupled BARWe model without exclusion 
\cite{Cardy-Tauber}. For stronger coupling strength ($s=2$) 
there are more blocking possibility resulting in 1-BARW2s 
scaling exponents.

In conclusion I have shown that the one dimensional 
two species coupled BARW with exclusion and
one offspring has the same critical transition point 
as that of the 2-BARW2s model investigated earlier. 
The hard-core interaction itself is not sufficient to 
cause deviation in scaling behaviour from that of DP.
A conjecture is given with regard the universality 
classes in coupled BARW systems exhibiting particle exclusion.

\vspace{3mm}
\noindent
{\bf Acknowledgements:}\\

The author would like to thank T. Antal for the 
stimulating discussions and N. Menyh\'ard for critically
reading the manuscript. Support from Hungarian research 
fund OTKA (Nos. T-25286 and T-23552) and from B\'olyai 
(No. BO/00142/99) is acknowledged.

\end{multicols}
\end{document}